# 3D Nanoporous Antennas for high sensitivity IR plasmonic sensing

Eugenio Calandrini[1], Giorgia Giovannini[1], and Denis Garoli[1,a)]

[1]*Istituto Italiano di Tecnologia, Via Morego 30, 16136 Genova, Italy*

a) denis.garoli@iit.it;

**Nanoporous gold can be exploited as plasmonic material for enhanced spectroscopy both in the visible and in the near infrared spectral regions. In particular, with respect to bulk metal it presents interesting optical properties in the infrared where it presents a significantly higher field confinement with respect to conventional materials. This latter can be exploited to achieve extremely high sensitivity to the environment conditions, hence realizing interesting sensors. Here we compare the sensitivity of a plasmonic resonators made of nanoporous gold with a similar structures made of bulk metal. The experimental test of the enhanced sensitivity was performed by depositing the same stoichiometric quantity of dielectric material onto the two considered structures. The result, also confirmed by the biosensing of a short peptide, can be ascribed to the better field confinement and enhancement in porous metal. This suggests an application of nanoporous 3D structures as sensor platform in the near-infrared with sensitivity over 4.000 nm/RIU.**

One of the most striking achievements pursued by plasmonics in optics is the enhancement of sensors sensitivity to detect molecular binding events and changes in molecular conformation [1–6,7]. The sensitivity of these sensors is therefore given by the ratio of the spectral shift $\Delta\lambda$ of the plasmon resonance due to the presence of the analyte and the change of the refractive index $\Delta n$ surrounding the plasmonic sensor. A significant dependence from the plasmon characteristics is hence expected. A key feature for characterizing the SPR sensor performances is the detection limit, or the minimum detectable amount of analyte, expressed in RIU. Several schemes have been successfully proposed, exploiting both the Surface Plasmon Polariton (SPP) and the Localized Surface Plasmon Resonance (LSPR) [8]. In the first case, propagating plasmons are launched on metal surfaces through the coupling with a prism[9]. The replacement of the bulky prism-coupling mechanism with ordered nanostructures like grooves or slits prepared on the metal surface provide the effective excitation of (SPP) at (near) normal incidence and allows the implementation of a microarray format for multiplexed and high-throughput biosensing platform[10]. Finally, their operation in transmission/reflectance configurations make them eligible for the integration in imaging-based devices [11]. On the other hand, nanoparticles with different sizes and shapes, either dispersed in colloidal solution[12] or fabricated on substrates by lithography[13,14], support localized plasmons. Plasmonic nanoparticles supporting LSPR displays greater spatial resolution, both lateral and normal, when compared with SPP and a scattering cross-section manifold greater than the fluorescence cross-section of fluorophores and they do not blink or bleach, providing a virtually unlimited photon resource for observing molecular binding over arbitrarily long time intervals. A complete review of LSPR-based sensors can be found in[15]. In both configurations, LSPR or SPP, plasmonic architectures feature subwavelength light confinement and enhancement near the metallic surface as a result of electromagnetic waves coupling to electrons. Since the resonance (and existence) conditions for these structures depend on the dielectric permittivity of the environment in contact to the surface of the metal, it turns out that plasmonics based detection is a refractometric measurement that responds to changes in the effective refractive index within the surface plasmon decay length. This latter dimension can spam between few to tens of nanometers depending on the working wavelength and on the structure design. Sensitivity of state-of-the-art plasmonic sensor typically peaks at 1000nm/RIU[16,17,18], and can even increases up to 30.000nm/RIU when plasmonic metamaterials are employed[19,20]. Unfortunately, this result was recorded exploiting a sensing mechanism based on the bulk Kretschmann configuration that is hardly miniaturized for commercial biosensing[21].

Recently we have shown that Nanoporous Gold (NPG) enables higher field confinement with respect to conventional bulk materials like gold and silver[22,23]. This is even more true in the near-infrared (NIR) spectral region and combined with the NPG's extensive surface (high surface/volume ratio) where an analytes can bind it makes the material extremely interesting for high sensitivity sensing. In particular we have recently demonstrated that the optical/plasmonic properties of NPG in the Near-IR can be exploited as a high performance sensing platform with sensitivity approaching 15.000 nm/RIU[24]. Even though higher sensitivities have been obtained by state-of-the-art platforms[19,20], this latter has been demonstrated in a very simple configuration where the spectral shifts of the NPG plasma edge were monitored during successive functionalizations with organic or dielectric molecules. Here we want to explore the NPG's sensing capability in the case of plasmonic resonators. In particular, considering our previous investigation on NPG 3D antennas[23], we build an SPR platform based on these structure for sensing in the NIR, and we compare its performance with the one obtained with a similar structure prepared in bulk gold. We will demonstrate a sensitivity over $4 \times 10^3$ nm/RIU by depositing conformal layer of $SiO_2$ on the NPG antennas. Moreover, we will report on detection of a small peptide (7-hystidine) at concentrations used in a previous work[24], still confirming the good performance of the platform.

Vertical antennas made of NPG have been fabricated following procedures described in details in Ref [23]. SEM micrographs of the prepared samples are reported in Figure 1 in comparison with the same structure prepared with bulk gold (in the inset). As well-known the SPR resonance wavelength depends on the length and the pitch of the antennas in the array. In our experiments we prepared vertical antennas with a height of 500 nm and a pitch of 1.25 μm respectively.

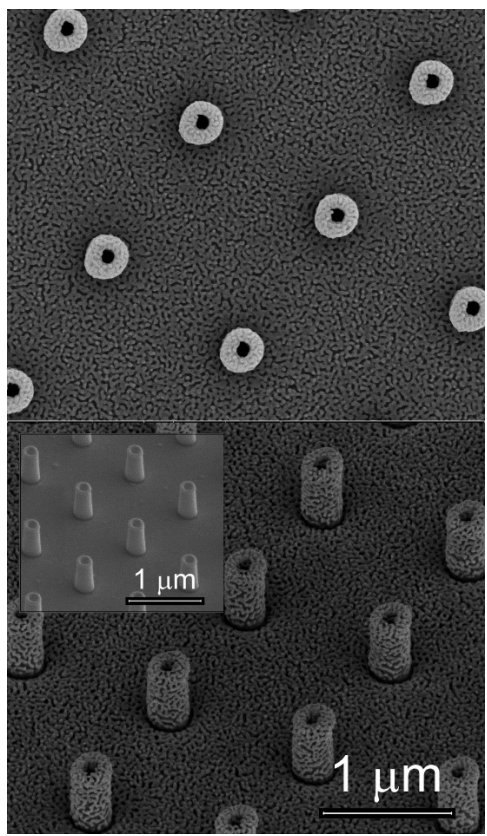

**Fig. 1.** SEM micrographs of NPG vertical antennas (top and tilted views). Inset: tilted view of the same structure prepared in bulk gold.

The bare structures have been characterized by means of FTIR and successively additional layers of $SiO_2$ have been deposited via atomic layer deposition (ALD) with steps of 1.4 nm putting the two samples (NPG and bulk) in the deposition chamber at the same time. As already mentioned, the intrinsic properties of ALD ensure stoichiometric depositions of the same amount of molecules onto the two substrates[25,26], hence allowing a direct comparison of the sensing performance without considering the higher surface / volume of NPG. The result of the measurement is illustrated in Figure 2(a) for both the NPG and the bulk gold, used here as a reference. The two antenna arrays were optimized to resonate at the same wavelength, within the fabrication accuracy. In this condition, the wavelength dependency of the refractive indexes of the employed materials - Au, NPG, $SiO_2$ - can be disregarded. Noteworthy, the NPG reflectance curves reports also the spectral shifts of the plasma edge. These are clearly more significant respect to the shift of the antenna's resonance. Following the minimum of the resonance dip as a function of the thickness of the $SiO_2$ layer, the spectral shift was calculated and plotted in Fig. 2(b) for the two arrays. The overall spectral shift for the NPG array is more than two-fold with respect the Au array and it is significant even for thicknesses around 1 nm. This remarks the fact that tuning the dielectric function of the plasmonic material is an additional degree of freedom to tailor for increase the sensitivity of plasmonic devices devoted to sensing applications.

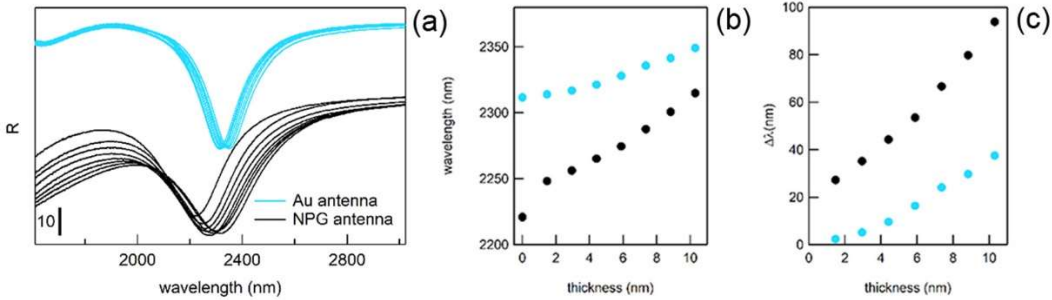

**Fig. 2.** (a)Reflectance curves of 3D antenna arrays for NPG and bulk gold. (b) Resonance wavelength and (c) spectral shift as a function of the thickness of the $SiO_2$ layer.

In order to evaluate the sensing performance of this system, the refractive index sensitivity S and the figure of merit FoM are introduced[27,28] . The first is defined by the nanometers of peak shift per refractive index unit and expressed in nm/RIU :

$S(t) = d\lambda(t) / dn(t)$     (1)

Where $d\lambda(t)$ is the spectral shift of the resonance peak as a function of the thickness of the deposited $SiO_2$ layer. The second evaluates its precision and depends on the ratio of sensitivity S and the peak linewidth $\Delta\lambda$

$FoM = S / \Delta\lambda$

While the spectral shift and the linewidth of the plasmonic resonance can be experimentally measured, for the refractive index shift an analytical analysis has been done following the criteria reported in Ref. [24]. It consists in treating the metal-$SiO_2$ material as Insulator-Metal multilayered metamaterial, whose overall effective dielectric function can be easily calculated applying the interface conditions for the electric field vector[29,30]. By using this approximation, it was possible to calculate the refractive index shift dn(t) induced by the depositions of $SiO_2$.

The results reported in Figure 3a show an average sensitivity for the NPG array of 4000 nm/RIU, whereas the Au array reports at best 1000 nm/RIU. Moreover, in the case of the first layers of the SiO$_2$ (1.4 nm), the NPG reports a sensitivity that peaks at 7000 nm/RIU, 17 times higher than Au. The performances of NPG array against the Au counterpart are confirmed by the FoM in Figure 3b. The FoM of ranges between 20 (calculated for $t_{SiO2}$ =10.5nm) and 30 (calculated for $t_{SiO2}$ =1.4nm) for NPG and it is 6 times higher than Au. Here it is important to note that although the NPG array peak linewidth is depleted, probably by its roughness, the FoM remains unaffected.

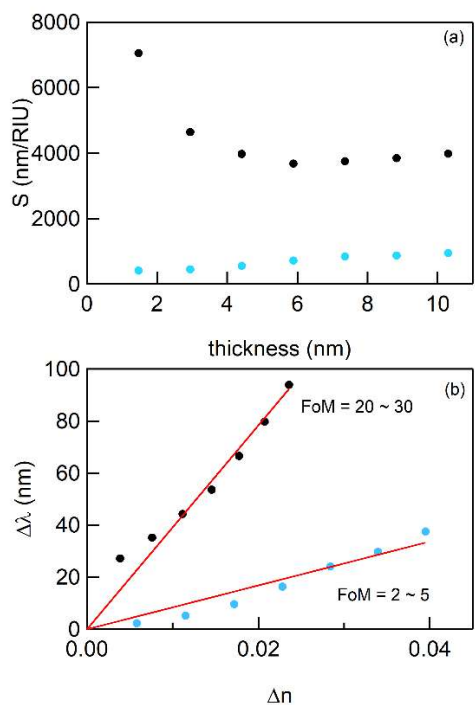

**Fig. 3.** (a) Sensitivity, expressed as nm/RIU, and (b) figures of merit of the NPG and Au arrays.

While the previous measurements can be used to evaluate the sensitivity of the platform, a real sensing application typically requires the detection of biomolecules at very low concentration. As in our previous experiments on NPG sensing[24], here we tested the performance of the material as a sensor by detecting low molecular weight proteins in solution. The detection approach was tested using a small molecule (c.a. 1kDa), in particular samples with a different concentration of polyhistidine (7HIS) were analyzed after the functionalization of the sensor surface (see methods). Obviously this approach can be successfully used also for the detection of bigger molecules such as proteins[31]. The covalent linkage of the analyte to the surface allows a better and more trustable detection, which can be accomplished both in wet and dry configurations. The results of the resonance peak shifts performed by means of FTIR for concentrations of 7HIS between 10pM and 10nM are illustrated in Figure 4. As in our previous experiments[24], by using this approach, a detection limit of 10 pM was verified, even though the spectral shift are significantly lower, in accordance with the lower sensitivity of the system. Noteworthy the ability of the NPG antennas platform to detect the functional layer of (3-Aminopropyl)triethoxysilane (APTES) deposited by means of ALD as conformal uniform monolayer. As previously proved[24], the thickness of this organic monolayer (measured by means of XPS) results about 0.6 nm. We expect that successive incubations of NPG 3D antennas with increasing concentrations of 7-HIS lead to increasing saturation of functional sites. The polypeptide has an expected thickness about 1.4 nm, hence, in total we expect an analyte layer 2 nm thick with a coverage that depends on the 7HIS concentration and the linkage efficiency. In the frequency region of the antenna resonance the refractive indexes of 7HIS and APTES are 1.70 and 1.47, respectively. At a final concentration of 10nM, we observe a spectral shift comparable to that obtained for $SiO_2$ deposition (extrapolating the shift expected with 2 nm).

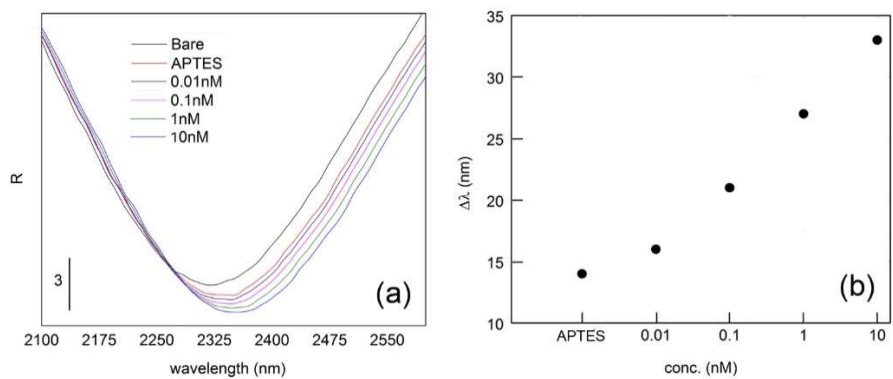

**Fig. 4.** Sensing performance of a bioanalyte. (a) Measured reflectance curves; (b) Spectral shifts.

The reported sensitivity can be explained by considering the plasmonic response of NPG in the NIR range. As illustrated in the authors' experiments, the field enhancement and confinement provided by the NPG at the metal/air or metal/liquid interface were much higher than what provided by the bulk counterpart[22]. This can be also a reasonable explanation for the observed higher sensitivity for the very first nm of functional layer (observed here and in our already mentioned work[24], both in the case of $SiO_2$ and APTES depositions). The porous structure, and the strong field confinement that rapidly decay above the very first nanometers of the NPG/air interface, enable a spontaneous co-localization of the plasmonic field and the analyte, thus promoting an effective interaction between them. This interaction results higher in the close proximity of the NPG surface, thus increasing the sensitivity to very small functional molecules / analytes.

In conclusion, here we shown that an interesting plasmonic material for IR sensing can be also exploited to prepare resonators which SPR shifts can be monitored to detected small amount of analytes with high sensitivity. The enhanced sensitivity obtained with the same stoichiometric amount of molecules on two different metals demonstrates that the effect is mainly due to a better field confinement and enhancement and not to the higher surface to volume ratio as previously proposed[32]. An easy and reproducibly fabrication strategy combined with the tunability of the optical properties[22] of the material confirm that nanoporous gold patterns appear promising for the realization of compact plasmonic platforms for sensing purposes. Compared to the bare film sensitivity of 15,000 nm/RIU reported in[24], here with an engineered nanostructure this quantity decreases down to 4,000 nm/RIU. This is probably because, approaching the $\omega_p$, the skin depth rapidly increases and the E field is able to probe the analyte molecule for all the film depth. On the other hand, an engineered nanostructure allows to target specific wavelengths.

METHODS

*Optical spectroscopy*

The optical response in the IR range was investigated by means of FT-IR reflectance microscopy measurements by using a Nicolet™ iS™ 50 FT-IR Spectrometer coupled to the Nicolet Continuμm Infrared Microscope by Thermo Scientific.™ A gold mirror was used as reference. The dielectric permittivity ε of this effective porous conductor was extracted through a fit of the measured reflectivity spectra within the Drude-Lorentz model. The optical response was characterized by a reduction of the absolute values of ε with respect to the bulk counterpart, mitigating the optical losses and increasing the skin depth, thus enabling the penetration of the electromagnetic field deeply into the porous matrix loaded with the analyte to be detected. FTIR microspectroscopy with a 30°-incidence Cassegrain objective in the reflection mode was employed to compare the far field optical response of two different chip of 3D vertical antennas made by nanoporous gold and bulk gold in a SPR configuration

*Refractive index calculation*

In the case of $SiO_2$ deposition, the refractive index variation $\Delta n$ was empirically calculated under the following assumption: the film covered by the $SiO_2$ can be thought of as Insulator-Metal multilayered metamaterial, whose overall effective dielectric function was dominated by the component longitudinal to the film plane. This can be easily calculated by applying the interface conditions for the electric field vector. The dielectric functions used for these estimations are found in literature for gold and silicon dioxide, while for NPG the previously calculated dielectric function was used. The complete analysis is reported in[24].

*Surface functionalization protocol for dry sensing experiments*

The NPG surface was firstly covered with APTES by ALD Self-limited deposition of single layer. The surface was then incubated for two hours at 40°C with 5mM (5mL) solution of succinic anhydride (>99%, Sigma Aldrich) in water. The surface was washed with water (5mL) to remove the unreacted anhydride, and the carboxyl groups thus exposed at the surface were activated with 5mM EDC (>99%, Sigma Aldrich) 5mM NHS(>98%, Sigma Aldrich) solution for 30 minutes at room temperature (R.T.). The coupling reagents were removed by flushing the surface with water. The surface was then treated with 10, 1, 0.1 and 0.01 nM solution of ployhistidine h at R.T. After 2 hours, the surface was washed once and analysed in dry condition.